\documentclass[final,5p,times,twocolumn,authoryear]{elsarticle}

\usepackage{graphicx} 
\usepackage{subcaption}
\usepackage{caption}
\usepackage{amsmath}
\usepackage{amssymb}
\usepackage{multirow}
\usepackage{comment}
\usepackage{xcolor, soul}
\usepackage{tabularx}
\usepackage{array}
\usepackage{booktabs}

\usepackage{hyperref}
\hypersetup{colorlinks = true, allcolors = cyan}
\usepackage[nameinlink,capitalise]{cleveref} 

\usepackage{tikz}
\usepackage{textcomp}
\usepackage{lipsum}

\newcommand\copyrighttext{%
  \footnotesize \textcopyright 2024 Elsevier. This article has been accepted for publication in Elsevier Computers \& Security and is made available under the  \href{https://creativecommons.org/licenses/by-nc-nd/4.0/}{CC-BY-NC-ND 4.0} license. This is the author's version which has not been fully edited and content may change prior to final publication. Citation information: DOI: \href{https://doi.org/10.1016/j.cose.2024.104144}{10.1016/j.cose.2024.104144}.
}

\newcommand\copyrightnotice{%
\begin{tikzpicture}[remember picture,overlay]
\node[anchor=south,yshift=760pt] at (current page.south) {\fbox{\parbox{\dimexpr\textwidth-\fboxsep-\fboxrule\relax}{\copyrighttext}}};
\end{tikzpicture}%
}

\renewcommand{\hl}[1]{#1}

\journal{Computers \& Security}

\begin{document}

\hypersetup{allcolors = cyan}

\begin{frontmatter}

\title{A Proactive Decoy Selection Scheme for Cyber Deception using MITRE ATT\&CK}

\author[inst]{Marco Zambianco\corref{corauthor}}
\ead{mzambianco@fbk.eu}
\author[inst]{Claudio Facchinetti}
\ead{cfacchinetti@fbk.eu}
\author[inst,inst2]{Domenico Siracusa}
\ead{domenico.siracusa@unitn.it}

\cortext[corauthor]{Corresponding author}
\affiliation[inst]{organization={Center for Cybersecurity, Fondazione Bruno Kesseler}, country={Italy}}
\affiliation[inst2]{organization={University of Trento}, country={Italy}}

\begin{abstract}
    Cyber deception allows compensating the late response of defenders countermeasures to the ever evolving tactics, techniques, and procedures (TTPs) of attackers.
    This proactive defense strategy employs decoys resembling legitimate system components to lure stealthy attackers within the defender environment, slowing and/or denying the accomplishment of their goals. 
    In this regard, the selection of decoys that can expose the techniques used by malicious users plays a central role to incentivize their engagement. However, this is a difficult task to achieve in practice, since it requires an accurate and realistic modeling of the attacker capabilities and his possible targets. In this work, we tackle this challenge and we design a decoy selection scheme that is supported by an adversarial modeling based on empirical observation of real-world attackers. We take advantage of a domain-specific threat modelling language using MITRE ATT\&CK\textsuperscript{\textcopyright} framework as source of attacker TTPs targeting enterprise systems. In detail, we extract the information about the execution preconditions of each technique as well as its possible effects on the environment to generate attack graphs modeling the adversary capabilities.
    Based on this, we formulate a graph partition problem that minimizes the number of decoys detecting a corresponding number of techniques employed in various attack paths directed to specific targets. 
    We compare our optimization-based decoy selection approach against several benchmark schemes that ignore the preconditions between the various attack steps. Results reveal that the proposed scheme provides the highest interception rate of attack paths using the lowest amount of decoys.
\end{abstract}


\end{frontmatter}

\copyrightnotice

\section{Introduction}\label{section:introduction}
In recent years, the enterprise sector has experienced an increasing number of cyber attacks targeting sensitive data exfiltration, IT infrastructure remote control, data destruction and/or encryption \citep{sharif2022literature}.
The involved threat actors leverage a plethora of tactics, techniques and procedures (TTPs) to gain a foothold within the production environment of an organization while remaining undetected, and then persistently achieving a variety of crimes over a long period of time.
In this scenario, the adoption of a proactive cyber defense mechanism, such as cyber deception, becomes pivotal to offset the strategic advantage of attackers over defenders by weaponizing the latter with decoys that mimic legitimate system components (e.g., text files containing fake login credentials, simulated communication protocols with vulnerabilities, fully emulation of
organization services) \citep{wang2018cyber}.  The allocation of these software artifacts within the production environment provides defenders the ability to reliably expose stealthy attackers whenever they engage these fake assets and, at the same time, the opportunity to monitor the employed TTPs without affecting the performance of the real system components.  
The information collected is extremely valuable for building reliable cyber threat intelligence that can be used to enhance the security posture of the system \citep{wagner2019cyber}.

Ideally,  an effective and practical cyber deception mechanism consists of decoys intentionally exposing vulnerabilities, misconfigurations and/or information that can be exploited by the attacker to progress towards his goal \citep{ferguson2021examining}. From a defender standpoint, the selection of decoys with these characteristics is a challenging task as it essentially requires a precise understating of the attacker behavior in order to anticipate his strategy. In this regard, the research activity in the field of cyber deception has partially addressed this problem \citep{sayed2023honeypot,cromp2023credential,kulkarni2020decoy,anwar2022honeypot}. 
In detail, attackers behavior are modeled as random walks in a graph where nodes identify organization assets that can be potential targets and edges indicate the feasible lateral movements of the attacker. Decoys are treated as abstract entities that are placed on non-target nodes to maximize the probability to intercept the attacker path according to some prior knowledge about the adversary attack steps. 
However, the proposed threat models are not sufficiently accurate to capture the attacker capabilities since they are not supported by empirical observations of the attacker TTPs. Moreover, the considered decoy modeling does not provide any practical insight on what weaknesses these artifacts should expose to reliably lure attackers before they can reach their targets. Consequently, both the applicability and effectiveness of such solutions in real world scenarios is limited. 

In light of these observations, we try to bridge the gap towards a more pragmatic cyber deception approach that is driven by an accurate modeling of TTPs employed by real-world attackers in enterprise environments. 
To accomplish this, we model attackers behavior using enterpriseLang  \citep{xiong2022cyber}, a threat modeling language based on the Meta Attack Language  \citep{johnson2018meta}. This domain-specific language provides a formal approach to describe adversaries using the MITRE ATT\&CK Enterprise matrix as source of knowledge of real-world TTPs targeting the IT infrastructure of enterprises \citep{mitre_attack}. This framework provides a record of cyber threats that are extracted from publicly available incident reports in the enterprise sector and that are organized by mapping similar threat actor strategies under the same techniques. The resulting threat model offers a holistic view of the attacker TTPs and related targets and it allows the generation of attack graphs that express the attacker capabilities in terms of techniques and related outcomes on the defender environment. Specifically, each node may correspond to either an ATT\&CK technique, whose predecessors indicate the required  ATT\&CK techniques enabling its execution, or to a technique outcome that reflects the impact of the attacker action on the defender environment. 

In the context of cyber deception, we take advantage of this threat model to anticipate adversarial behavior and proactively select decoys that can be engaged by the attacker using specific ATT\&CK techniques.
For example, a text file decoy is suitable to lure adversaries employing the ATT\&CK technique
“Remote System Discovery” which targets the collection of the system elements information from local host files. In this case, a defender that wants to expose an adversary using this technique could 
locate fake hostnames in a text file located in a monitored system folder that exclusively non-legitimate users would try to access. By extending this reasoning to every ATT\&CK techniques, 
we optimize the selection of decoys by computing the minimum subset of techniques that, if detected by a corresponding number of deceptive artifacts, ensure the interception of all attack paths executed by an attacker to reach a predetermined set of targets. In detail, we resume our contributions as follows:

\begin{itemize}
    \item We propose a methodology to proactively select decoys in enterprise environments using the MITRE ATT\&CK Enterprise matrix. 
    In particular, we leverage the cyber threat information provided by the threat modeling language enterpriseLang to coherently combine ATT\&CK techniques according to their execution preconditions. We exploit this data-driven attack path modeling to design a decoy selection scheme tailored to realistic attacker engagement policies.
    \item We analytically model the decoy selection problem as a graph partition problem. Specifically, we formulate a linear integer problem, based on the Multi-Terminal Vertex Separator Problem, to compute the smallest subset of nodes (i.e., attacker techniques) in the attack graph that disconnects a set of source nodes (i.e., the attacker initial capabilities) from a predetermined set of target nodes, representing the attacker targets. As a result, the deployment of decoys exposing the selected techniques can intercept all possible attack paths.
    \item We evaluate the performance of our approach designing several benchmark schemes and metrics. The proposed decoy selection approach ensures the interception of all possible attack paths using the lowest amount of decoys regardless of the number of attack targets. In particular, thanks to the employment of a threat model that coherently combines the ATT\&CK techniques based on their execution preconditions, our solution outperforms an approach that selects decoys according to ATT\&CK techniques considered in isolation from one another.
\end{itemize}

The remainder of the paper is organized as follows. In \cref{section:related_work}, we discuss the related work. In \cref{section:background}, we present some background information. In \cref{section:system_model}, we describe the considered system model, which is composed by the threat and defender models. In \cref{section:problem_formulation}, we formalize the proposed optimization problem that computes the selection of decoys. In \cref{section:performance_evaluation}, we present the simulation results. Finally, we draw the conclusion in Section \cref{section:conclusion}.

\section{Related work}\label{section:related_work}

\begin{table*}[t]
    \caption{\hl{Comparison of cyber deception solutions}}
    \centering
    \footnotesize
    \begin{tabularx}{0.9\textwidth}{@{}p{3.5cm}p{2.0cm}p{2.4cm}p{3.0cm}p{3.5cm}@{}}
        \toprule
        \textbf{Reference} & \textbf{Focus} & \textbf{Threat Model} & \textbf{Objective}   & \textbf{Limitations} \\ 
        \midrule
        \cite{horak2019optimizing,al2021hidden,nguemkam2023optimal} & Decoy allocation  &  Lateral movement & Allocation of decoys to maximize the attacker interaction probability & Threat model ignores specific TTPs; decoys are not practically characterized \\ 
        \midrule
        \cite{alohaly2022integrating,pagnotta2023dolos} & Decoy implementation & Lateral movement & Lightweight decoy implementation integrated within production environment to detect insider threats & Decoy design is not driven by a quantitative modeling of attacker TTPs\\
        \midrule
        \cite{ge2021proactive,rehman2024proactive} & Decoy allocation & Reconnaissance, Data exfiltration & Allocation of decoys to distort attacker perception of IoT environments & Consider only specific attack stages, hence they cannot be extended to more complex attack scenarios composed by multiple attack steps \\ 
        \midrule
        \cite{brown2023evaluating,sajid2021soda,mashima2022mitre,subhan2023unveiling,hobert2023enhancing} & Decoy selection  & ATT\&CK techniques & Selection of decoys capable to detect specific MITRE ATT\&CK techniques & Selected decoys ignore possible defense mitigation already in place; techniques preconditions are not exploited to simultaneously intercept multiple attack paths  \\ 
        \midrule
    \end{tabularx}
    \label{tab:sota}
\end{table*}

Recent work in cyber deception has evolved towards the design of decoys, honeypots and /or honey-tokens  solutions to lure TTPs targeting specific defender environments such as enterprise, industrial, internet-of-things. An overview of the proposed solutions and related threat models can be found in \citep{qin2024hybrid,zhang2021three}. Moreover, an increasing number of deception approaches leverage cyber threat intelligence platforms such as MITRE ATT\&CK to design and/or evaluate the proposed schemes, ensuring more effective and reliable technologies for practical scenarios \citep{torres2022cyber}. In this context, we highlight our contributions by emphasizing the difference in the methodology that we propose in building a threat model supported by real-world adversary behavior that also captures the execution preconditions between the various techniques. In \Cref{tab:sota}, \hl{we provide a state-of-the-art summary of the considered works.}

The authors of  \citep{horak2019optimizing} and \citep{al2021hidden} present decoy allocation schemes based on hidden Markov models and partially observable scholastic games, respectively, to predict the most probable attack path of an adversary and place the deceptive artifacts accordingly. 
The authors of \citep{nguemkam2023optimal} propose a honeypot allocation scheme that leverages core attack graphs to reduce the number of computed attack paths with limited impact on the deception performance, hence improving deployment scalability. 
Although these works design practical approximations to render the related algorithms efficient in practical scenarios, the proposed threat models are limited to laterally movements between the various system elements and they ignore the specific TTPs used by attackers to interact and penetrate those components. As a result, they do not provide information about the functionalities that each decoy should implement in order to lure a malicious user. Instead, we leverage a threat modeling language to  extend the attacker capabilities beyond abstract lateral movements among system assets using real-world TTPs. This strategy allows us to select decoys tailored to detect specific techniques that implicitly offer practical insights about the deceptive artifact implementation.

The authors of \citep{alohaly2022integrating} and \citep{pagnotta2023dolos} propose a lightweight implementation of deception solutions to detect insider threats within production environments of enterprise systems. Although, their approach is orthogonal to our as we focus on the decoy selection rather than its technical implementation, the lack of a quantitative threat modeling of the attacker behavior prevents a proper performance evaluation, hindering their effectiveness.  Differently, we consider a threat model generating attack paths that are constituted by techniques employed in real enterprise environments. This approach enables a proactive decoy allocation, where the defender can design a practical deception strategy anticipating the adversary behavior. 

The authors of \citep{ge2021proactive} and \citep{rehman2024proactive} propose several cyber deception strategies based on different methodologies including genetic optimization algorithms to distort attacker perception of IoT environments using decoys. However, they tailor their decoy allocation to reconnaissance and data exfiltration attacks only. Moreover, they do not  model the intermediate attack steps to achieve those goals. Consequently, the considered threat model is neither suitable for attack scenarios in the enterprise environments nor it can provide a quantitative characterization of the attacker behavior due to the limited number of considered attack paths.

The authors of \citep{brown2023evaluating} propose MTDSim, an framework to evaluate the performance of Moving Target Defense and cyber deception approaches against realistic attackers.  The threat model is derived using ATT\&CK tactics only, hence it considers the various phases of a cyberattack without providing information about each attack step. Instead, we provide a representation of the attacker behavior by extracting the execution preconditions between techniques from a threat language model using MITRE ATT\&CK as TTPs source. This enables the optimization of the amount of selected decoys as we can exploit the execution preconditions between techniques  to minimize the number of deceptive artifacts.

In \citep{sajid2021soda}, the authors propose SODA, an orchestration framework that automatically analyzes real-world malware behavior, maps its activities to ATT\&CK techniques and design suitable deceptive actions to counteract any malicious software performing those operations on the real system. 
The authors of \citep{mashima2022mitre} provide a performance evaluation of a cyber deception solution for smart grid systems using MITRE ATT\&CK. In particular, they map real-world attack paths to ATT\&CK techniques and they empirically show that the considered decoy architecture offers a mitigation to each attack step. Differently, the authors of \citep{subhan2023unveiling} and \citep{hobert2023enhancing} collect cyber threat intelligence data from deployed honeypots and map the observed attacks to ATT\&CK techniques. This information is used to provide a starting point for possible technique mitigations and to cluster similar approaches under a single threat actor, respectively.
The aforementioned work employ MITRE ATT\&CK to evaluate a-posteriori the effectiveness  of the proposed deception strategies in terms of detected techniques. We propose a fundamentally different approach that exploits MITRE ATT\&CK to generate attack graphs modeling the various attack steps of an adversary. Based on this, we design the decoy selection to expose the minimum number of techniques used by all feasible attack paths aiming to a predetermined set of targets.

\section{Background}\label{section:background}

We summarize the main concepts of MITRE ATT\&CK framework and EnterpriseLang threat modeling language using the respective original works \citep{mitre_attack,xiong2022cyber}.
We remark that these tools are the core elements that we employed to design our decoy selection solution and their description is helpful to provide a better understanding of our contribution in later sections. 

\subsection{MITRE ATT\&CK framework overview}
The MITRE ATT\&CK Framework provides a structured classification, inspired by the cyber kill chain model \citep{hutchins2011intelligence}, of real-world cyber attacks mainly targeting enterprise IT infrastructures. 
At a high level, ATT\&CK can be considered as an adversary behavioral model that represents observed attacks according to the following nomenclature:
\begin{itemize}
    \item \textbf{Tactics}: they represent the short-term tactical goal of an adversary and essentially correspond to the reason for performing a technique. For example, an adversary wants to achieve \textit{Credential Access} as an intermediate goal in order to accomplish the final target that may be data exfiltration.
    \item \textbf{Techniques}: they represent how the adversary achieves its tactical goal by describing the general strategy to accomplish that objective. For example, an adversary aiming to achieve  \textit{Credential Access} may search for typical password storage locations within the victim system to obtain the user credential. Each technique may be further split in sub-techniques when several existing strategies are available.

\end{itemize}
Threat actors showing similar behavior are clustered and considered as a single adversary, referred to as \textit{Group}. Their activities are mapped according to the above nomenclature in a matrix where columns are the tactics and rows are the techniques. For each technique, the associated mitigations are also present (when available). As a result, by inspecting a particular group, it is possible to examine how those threat actors operated in the past to achieve specific targets. 

\subsection{EnterpriseLang overview}
EnterpriseLang is a threat modeling language that enables the formal description of the adversary behavior in enterprise systems. Its structure is based on the Meta Attack Language (MAL), a general-purpose threat model language that provides  syntax and symbol notation to create attacker models for specific domains (e.g., cloud, automotive, internet of things, etc.). For the enterprise scenario, enterpriseLang models the attacker actions using the cyber threat intelligence information provided by the MITRE ATT\&CK enterprise matrix. 
 Each one of the 266 ATT\&CK technique is translated according to the MAL specification in the following items, which are the main elements constituting the language architecture: 
 \begin{itemize}
    \item \textbf{Attack Step}: it represents the action of the adversary and is identified by an ATT\&CK technique with the related outcome on the defender environment. Each outcome characterizes the effects of the various ATT\&CK techniques on the environment assets (note that some techniques may not have an associated outcome since they do not have an immediate impact on the environment). Examples of technique outcomes include \textit{SSH Credential Interception}, \textit{Remote Execution}, \textit{Sensitive Data Collected}, \textit{Infected Computer}. 
     \item \textbf{Connections}: they link the execution preconditions needed by the adversary in order to take advantage of a technique. In practice, the various connections indicate all the required attack steps that must be executed by the attacker in order to employ the considered technique.  
     \item \textbf{Type}: it indicates when an adversary is eligible to execute an attack step based on its connections. There are two possible type categories, namely AND type and OR type.  The former identifies an attack step that can be processed by the attacker only if every predecessor attack step is successfully completed. Conversely, the latter identifies an attack step that can be executed as soon as at least one of its predecessor attack step is finalized.
     \item \textbf{Defense mitigations}: they correspond to possible countermeasures, if available, that can be adopted by the defender to prevent the working of a technique and its associated outcome. Example of mitigations can range from enforcement of security policies to the update of software having known vulnerabilities.
 \end{itemize}

Leveraging the information encoded among the various language elements, we express attack paths in an enterprise environment by coherently combining multiple ATT\&CK techniques according to their cause-effect relationship.

\section{System Model}\label{section:system_model}

\begin{figure*}[t]
    \centering
    \includegraphics[trim={3cm 3.5cm 3cm 3cm},clip,width=0.8\linewidth]{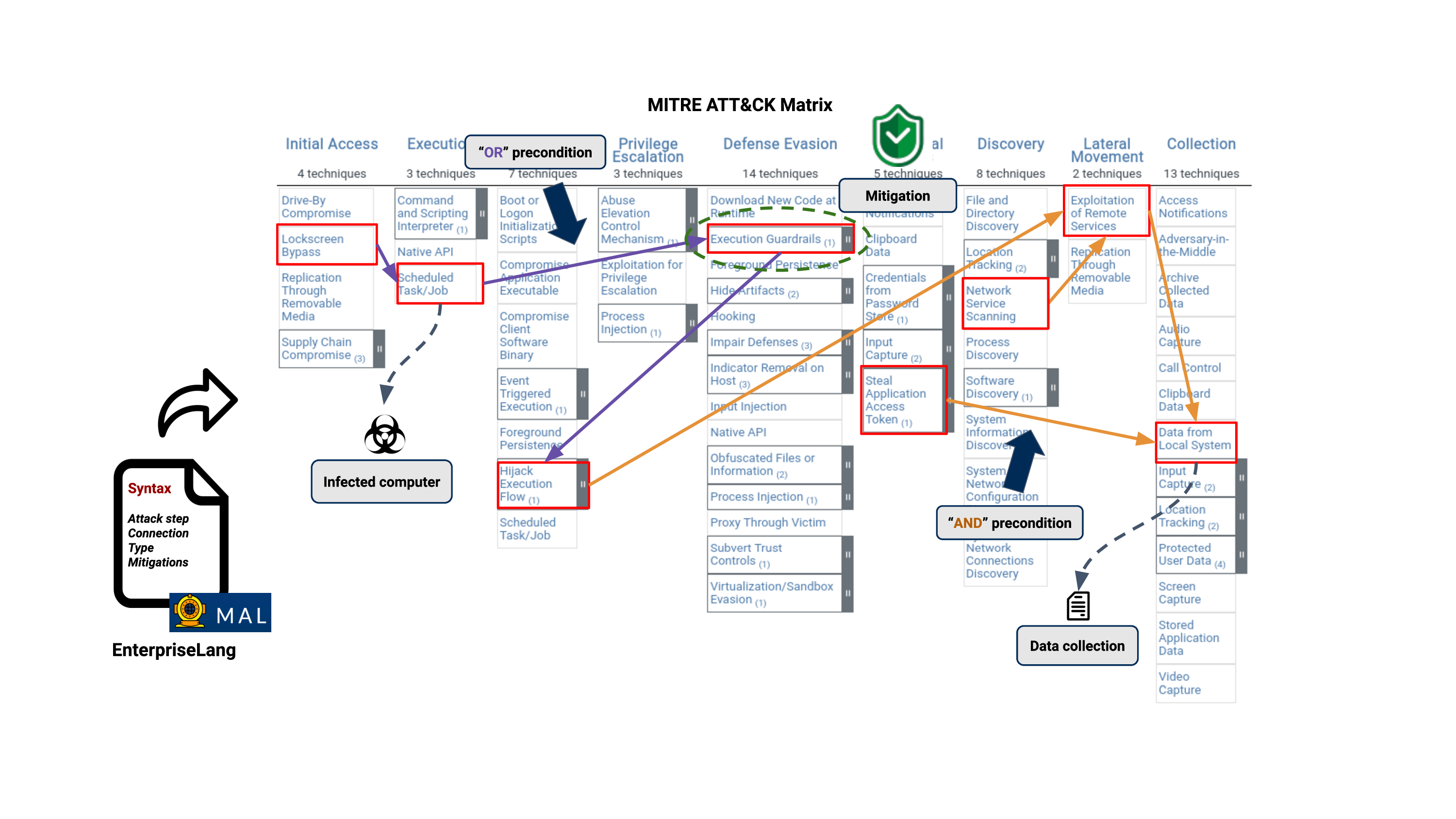}
    \caption{\hl{Visual representation of the proposed threat model. The attacker capabilities are expressed as a subset of ATT\&CK techniques (red squares) of the MITRE ATT\&CK Enterprise matrix, where techniques within the same column belong to the same tactic. The considered techniques are combined according to their execution preconditions using \textit{Connection} and \textit{Type} information encoded within enterpriseLang. Similarly, the attack path representation is further augmented with the technique outcomes and possible defense mitigations using  \textit{Attack Step} and \textit{Defense mitigations} information, respectively.}}
    \label{Fig:threat_model}
\end{figure*}

We consider an organization (i.e., the defender) that employs cyber deception to reliably detect attackers aiming to some targets within his IT infrastructure. To ensure an effective decoy placement that is tailored to real-world attacker capabilities, it leverages enterpriseLang to model the adversary behavior in terms of ATT\&CK techniques and uses such representation to design the decoy selection process. In the next subsections, we formalize the threat model considered and how it is used by the defender to achieve a practical decoy selection solution. We resume the main notation used in the paper in \Cref{tab:glossary}.

\subsection{Threat Model}

We define $A$ as  the set of ATT\&CK techniques provided by the enterpriseLang, where we further define various techniques subsets according to their \textit{Type} and \textit{Defense mitigations} information (see the background section for their description) as follows:
\begin{itemize}
    \item $A^M$ is the subset of techniques with defense mitigations available.
    \item $A^{\bar{M}}$ is the subset of techniques without defense mitigations available.
    \item $A^{AND}$ is the subset of techniques with preconditions of type $AND$.
    \item $A^{OR}$ is the subset of techniques with preconditions of type $OR$.
\end{itemize}
From the above definition, we remark that $A = A^M \cup A^{\bar{M}}$ and $A = A^{AND} \cup A^{OR}$, hence each technique may have a mitigation and its execution preconditions can be either of type $AND$ or type $OR$. Following a similar reasoning, we define $O$ as the set of technique outcomes caused by the execution of the various techniques, where we further define the subsets $O^{AND}$ and $O^{OR}$ to indicate outcomes of type \textit{AND} and type \textit{OR}, respectively. There are no mitigations defined for the outcomes since they are a consequence of the execution of a technique, hence an active defensive mitigation for that technique would automatically prevent its outcome. 

We formally represent the attack graph by combining the element of $A$ and $O$ in a directed graph $G=(V, E)$. Specifically, $V = A \cup O$ is  the set of nodes where each element $v_i \in V$  corresponds to a technique if $i \in A$ or to a target if $i \in O$. Similarly, we define $E = \{(v_i,v_j) \subseteq V \times V\} $ as the set of edges which connect the various nodes according to their relationship:  
\begin{itemize}
    \item If $v_i, i \in A$ and $v_j, j \in A$ the edge $(v_i,v_j)$ indicates that technique $i$ is a precondition for the execution of technique $j$. For example, the technique \textit{Hijack Execution Flow}, which enables to run malicious code by manipulating how the operating system runs programs, must be first executed by the attacker to employ the  technique \textit{OS Credential Dumping}, which allows the location and theft of credentials material from the operating system and software.
    \item If $v_i, i \in A$ and $v_j, j \in O$ the edge $(v_i,v_j)$ indicates that technique $i$ produces the technique outcome $j$ on the defender environment. For example, the execution of the technique \textit{Application Shimming}, which enables to run malicious code triggered by application shims, causes the outcome \textit{Bypass User Access Control}, which allows the attacker to bypass the security feature on Windows systems restricting  the execution of software based on specific user privileges.
    \item If $v_i, i \in O$ and $v_j, j \in A$ the edge $(v_i,v_j)$ indicates that the technique outcome $i$ enables the attacker to employ technique $j$.  For example, the technique outcome \textit{Infected Computer} enables the adversary to accomplish a variety of techniques like \textit{Data Encrypted}, \textit{Data Destruction}, \textit{Account Manipulation}.
\end{itemize}

The resulting attack graph can be considered as an augmented version of the original MITRE ATT\&CK Enterprise matrix that models the dependency between each technique in terms of execution preconditions as well as their impact on the target environment. In detail, we can assess how an attacker can practically achieve the execution of a technique in the environment by inspecting the predecessor nodes in the graph. The same reasoning applies when analyzing how an outcome can be achieved starting from a technique. In \cref{Fig:threat_model}, \hl{we show a pictorial overview of how we combine the information provided by the MITRE ATT\&CK framework and enterpriseLang to build the proposed threat model.} 

\begin{figure}
    \centering
    \includegraphics[trim={2cm 2.5cm 2cm 3cm},scale=0.4,clip]{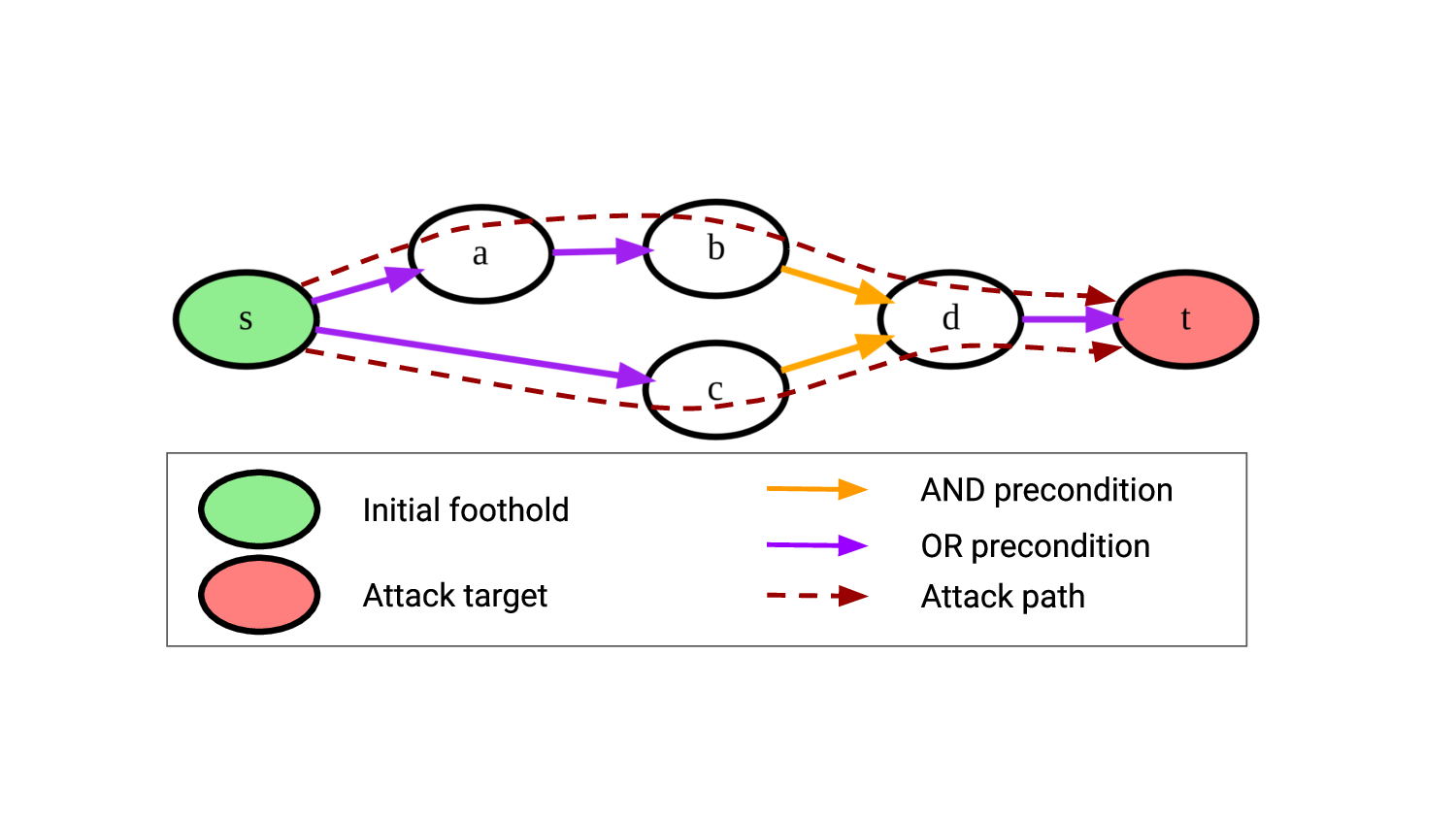}
    \caption{Example of attack path when AND nodes are included. Purple arrows represent OR preconditions. Orange arrows represent AND preconditions. Red dashed arrows indicate the attack path from node $v_s$ to node $v_t$.}
    \label{Fig:paths_example}
\end{figure}

To model the representation of such cyber-threat intelligence information, we define the concept of attack path as a simple directed path (i.e., a path containing no repeated vertices) between a node pair $(v_s,v_t),\forall s,t \in A \cup O$, where $v_s$ is the source node and $v_t$ is the target node. In addition, if an attack path contains nodes of type AND, all the related predecessors (which must be reachable from the source node $v_s$) are also included in order to correctly describe the execution preconditions of that technique. Note that there can be multiple attack paths for the same source-target node pair as any directed path reaching $v_t$ from $v_s$ may be a candidate attack path if all nodes of type \textit{AND} within the path can be reached as well. In this context, we can formalize the notion of directed simple path as $P_{s,t} = \{(v_{s=1},v_{2},...,v_{t=|P_{s,t}|}) : (v_i,v_{i+1}) \in E, i=1,...,|P_{s,t}|\}$. 
Leveraging this definition, we analytically express an attack path as
\begin{multline}
        AP(s,t) = P_{s,t} \cup \{N_G(v_i), \forall v_i \in P_{s,t} : \\  i \in A^{AND} \cup O^{AND} \land N_G(v_i) \subseteq R_G(v_s) \},
        \label{eq:ap_def}
\end{multline}

where $N_G(i)$ indicates the set of predecessor nodes connected to node $v_i$ and $R_G(v_s)$ indicates the set of all nodes that can be reached from node $v_s$. 
In other words, an attack path is identified by i) the nodes composing a simple directed path starting from source node $v_s$ to target node $v_t$ and by ii) the predecessor nodes of the \textit{AND} nodes, contained in the original simple directed path, that are reachable from the source node $v_s$. Therefore, attack paths composed by any number of \textit{AND} nodes can be seen as a bundle of simple directed paths connecting node $v_s$ to node $v_s$, since each \textit{AND} node generates a shorter attack path between $v_s$ and its predecessors. We provide an illustrative example of this behavior to better visualize this dynamic in \cref{Fig:paths_example}. In this case, there is a single \textit{AND} node with $2$ predecessors in a directed simple path, hence the actual attack path is composed by at least $2$ parallel simple paths, identified by the dashed line, as every technique associated to predecessor nodes must be executed by the attacker.

\begin{table}[t]
    \centering
    \scriptsize
    \caption{System notation}
    \vspace{-0.2cm}
    \begin{tabular}{ |l|l|  }
        \hline
        $A$ & Set of ATT\&CK techniques \\
        \hline
        $O$ & Set of technique outcomes \\
        \hline
        $V$ & Set of vertices of attack graph $G$ \\
        \hline
        $E$ & Set of directed edges of attack graph $G$ \\
        \hline
        $N_G(v_i)$ & Set of predecessor nodes of $v_i$ in $G$ \\
        \hline
        $S$ & Set of source nodes \\
        \hline
        $T$ & Set of targets nodes \\
        \hline
        $R_G(v_i)$ & Set of reachable nodes from node $v_i$ in $G$ \\
        \hline
        $P_{s,t}$ & Simple directed path from  node $v_s$ to $v_t$ \\
        \hline
        $AP(s, t)$ & Attack path from node $v_s$ to $v_t$ \\
        \hline
        \(\bar{G}\) & Subgraph induced by attack paths from $S$ to $T$ \\
        \hline

    \end{tabular}
    \label{tab:glossary}
\end{table}

\subsection{Defender Model}

To implement a proactive cyber deception approach,  the defender leverages the proposed attack graph to model the expected adversary behavior by means of attack paths, that indeed express which ATT\&CK techniques are required and how they are correlated with one another to accomplish some goal. In detail, we assume that the defender is unaware of the attacker goals, hence he selects a set of technique outcomes $T \subseteq O$ that would severely damage the system once accomplished and considers them as attack targets of a hypothetical adversary. Based on this selection, the defender analyzes which attack steps are required to reach the various targets from a set of techniques and/or outcomes $S = \{s:  s \in A \cup O \land  s \notin T \}$, that model the attacker initial capabilities. 
The selection of the elements in $S$ can follow different criteria. For example, the defender could be interested in tracking the adversary progression towards his goals starting from different attack stages. Using as reference \Cref{Fig:paths_example}, node $v_s$ could correspond to the technique \textit{Exploitation of Remote Services} that indicates the capability of an attacker to move laterally between internal system components of the defender environment exploiting his production services. This scenario indicates an attacker that has already gained persistence within the system by evading security countermeasures. The attack paths originating from such node are likely to employ more advanced attack tactics such as \textit{Command and Control} and \textit{Exfiltration}.  Differently, the analysis of attack paths including early attack stages could be indicated by configuring node $v_s$ with the technique \textit{Shared Module}. The latter indicates an adversary expanding his initial foothold within the system by exploiting shared executable files between different process to run arbitrary code.

For each source and target pair, we compute each attack path according to \eqref{eq:ap_def}. We comprehensively represent all paths by extracting the induced subgraph $\bar{G} = (\bar{V},\bar{E})$, also referred to as \textit{threat profile}, where:
\begin{itemize}
    \item $\bar{V} = \left\{\bigcup_{} AP(i,j), \forall (v_i,v_j) \in S \times T \right\}$ is the set of vertices and corresponds to the union of all techniques and outcomes in every attack path between the various source-target node pair $(v_i,v_j)$ , $v_i \in S, v_j \in T$.
    \item $\bar{E} = \{(v_i,v_j) \subseteq \bar{V} \times \bar{V}\}$ is the set of edges that connect each node pair $(v_i,v_j)$, as in the original attack graph $G$, when both $v_i, v_j \in \bar{V}$. 
\end{itemize}

The defender employs the resulting threat profile to optimize the selection of decoys to deploy within its system. In particular, we assume that the defender has the availability to deploy decoys that are specifically tailored to expose an attacker using each technique in the threat profile. In other words, for every ATT\&CK technique, there exists a decoy suited for its detection. Our assumption is supported by MITRE frameworks, namely Engage \citep{mitre_engage} and Def3nd \citep{mitre_defend}, that offer tools and resources to increase the effectiveness of adversary engagement strategies like cyber deception. In particular, they provide a mapping between ATT\&CK techniques and possible decoys in order to facilitate the design of cyber deception solutions for cybersecurity practitioners. Although this mapping is not exhaustive for all ATT\&CK techniques, we consider that it exists for any technique to simplify the design of the proposed scheme. This is still a reasonable assumption as any software artifact that can mimic a functionality of the legitimate system can serve as decoy. Consequently, any ATT\&CK technique that has been observed to target or leverage those system components can be exposed by a suitably crafted decoy. Nonetheless, we acknowledge that this reasoning neglects the actual implementation feasibility required to create such decoys that might be prohibitive for real-world scenarios. For this reason, we leave as a future analysis the investigation of a decoy selection solution where only a limited subset of the adversary techniques can be exposed by a decoy.

In addition to the primary goal of ensuring a reliable detection of the attacker, the defender designs the decoy selection strategy to meet two main requirements:
\begin{itemize}
    \item \textbf{Attack surface minimization}: the number of allocated decoys is minimized to limit the attack surface increase. The introduction of any new component (either software or hardware) in an IT infrastructure  can be potentially exploited by an attacker to further compromise the environment. Therefore, it is a good cybersecurity practice to generally restrict the injection of new elements to limit the attack surface increase \citep{theisen2018attack}. This reasoning still applies to decoys that may be possibly intertwined with the production environment and thus could be exploited unpredictably to damage legitimate system components.
    \item \textbf{Unmitigated techniques maximization}: the allocated decoys prioritizes the exposure of techniques without available mitigations. This approach accounts for the presence of different security countermeasures within the same system and offers the opportunity to compensate possible security gaps that cannot be filled using traditional cybersecurity approaches (e.g. firewall, antivirus, intrusion detection tools, etc.).
\end{itemize}
In the next section, we present an optimization-based decoy selection scheme that allows to effectively satisfy the aforementioned requirements while ensuring a reliable adversary interception performance based on the considered threat profile.

\section{Problem formulation}\label{section:problem_formulation}

\begin{figure*}[t]
    \centering
    \includegraphics[width=0.8\linewidth]{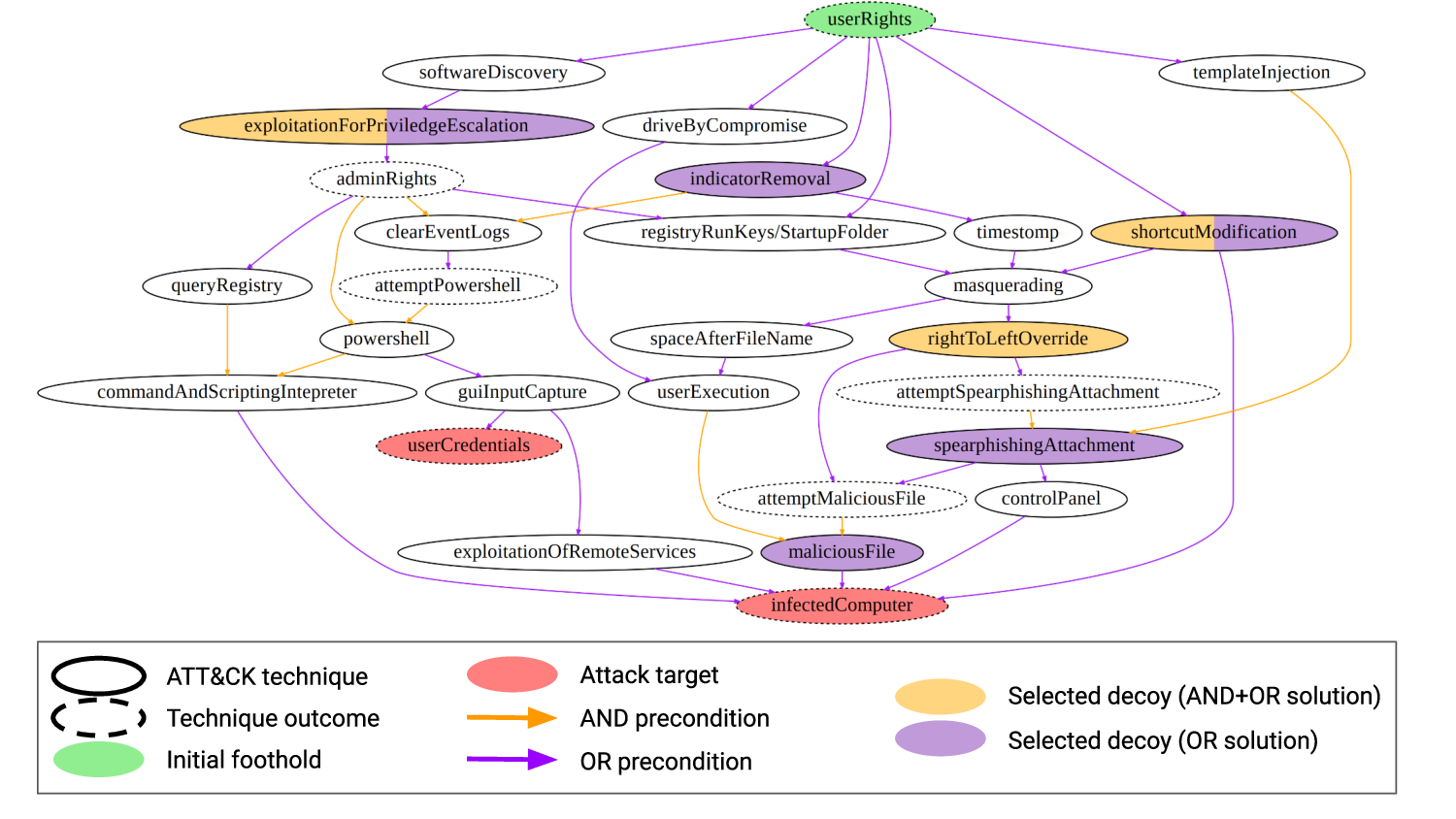}
    \caption{\hl{Example of decoy selection solutions when AND nodes and OR nodes are both considered (AND+OR solution, yellow nodes) and when  OR nodes only are considered (OR solution, purple nodes).}}
    \label{fig:mal_example}
\end{figure*}

The problem of minimizing the number of selected decoys for a given threat profile $\bar{G}$ can be analytically model as graph partition problem. In detail, we compute the minimum number of nodes (i.e., techniques) that makes each node in $T$ unreachable from $S$, hence denying any possible attack path. Consequently, the deployment of decoys exposing the selected techniques can reliably intercept an attacker that has an initial foothold within the defender environment expressed by techniques and/or outcomes in $S$ and that aims at the targets in $T$. In the next subsections, we describe the problem formulation as well as an example of solution computation.

\subsection{Optimal decoy selection}
We analytically compute the selection of deceptive artifacts by proposing an integer optimization problem that can be considered as a variant of the Multi-Terminal Vertex Separator problem \citep{MTVSP}.  The general idea is to partition the graph vertices into three different subsets $X, Y$ and $Z$ such that vertices in $X$, if removed, disconnect all paths from any vertex in $Y$ to any vertex in $Z$. We indicate the node assignment to a partition by defining the binary indicator variables $x_i, y_i, z_i$ that individually assume value $1$ if vertex $v_i \in X$, $v_i \in Y$, $v_i \in Z$, respectively, and value 0 otherwise.

The set $Y$ and $Z$ represent the source and target partitions and are initialized by setting $y_i = 1, \forall v_i \in S$ and $z_i = 1, \forall v_i \in T $, respectively. Differently, partition $X$ is initialized as an empty set and represents the decoys exposing techniques disconnecting $Y$ from $Z$. Every vertex must be assigned to exactly one partition, hence all the unnecessary vertices not belonging to $X$ are either assigned to  $Y$ or $Z$. To correctly compute the partition $X$, we need to account for the type of every node.  Nodes of type \textit{AND} impose the reachability of all their predecessors in order to be employed in the attack paths, whereas nodes of type \textit{OR} require the reachability of at least one of their predecessors. We encode this information by further defining the binary indicator variable $r_{i,s}$ that assumes value 1 if node $v_i \in R_{\bar{G}}(v_s), \forall v_s \in S$ (i.e., if node $v_i$ is reachable from any node $v_s \in S$). 
Finally, we introduce the variable $c_i$ that assume a fixed value of $1$ if $i \in A^{\bar{M}}$ and variable value $\beta \geq 1$ if $i \in A^{M}$. This variable allows associating a higher cost to techniques with mitigations in order to disincentivize their selection during the computation of the partition $X$. Leveraging the introduced variables, the optimization problem can be written as

\begin{equation}
\min_{\boldsymbol{x}} \quad  \sum\limits_{v_i \in A} c_i \cdot x_i
\label{eq:obj_func_new}
\end{equation}

subject to

\begin{align}
\label{eq:reachable_or}
r_{i,s} =  \min \{ y_i + z_i,  & \sum_{n \in N_{\bar{G}}(v_i)} r_{n,s} \} \\ 
& \forall v_i \in A^{OR} \cup O^{OR}, \forall v_s \in S \nonumber
\end{align}

\begin{align}
\label{eq:reachable_and}
r_{i,s} =  1 - \min \{ 1, \quad x_i + & \sum_{n \in N_{\bar{G}}(v_i)}  1 - r_{n,s} \} \\ 
& \forall v_i \in A^{AND}  \cup  O^{AND}, \forall v_s \in S \nonumber
\end{align}

\begin{equation}
y_i + z_j \leq 1 + ( 1 - r_{i, s} ) + ( 1 - r_{j, s} ) \quad \forall (v_i, v_j) \in \bar{E}, \forall v_s \in S
\label{eq:cross_boundaries}
\end{equation}

\begin{equation}
y_s = 1, z_t = 1, x_i = 0 \quad \forall v_s \in S, \forall v_t \in T, \forall v_i \in O 
\label{eq:init_var}
\end{equation}

\begin{equation}
x_i + y_i + z_i = 1 \quad \forall v_i \in \bar{V}
\label{eq:one_cluster}
\end{equation}

\begin{equation}
x_i, y_i, z_i, r_{i, s} \in \{0,1\} \quad \forall v_i,v_s \in \bar{V}
\label{eq:binary}
\end{equation}

The objective function \eqref{eq:obj_func_new} minimizes the total number of nodes belonging to $X$, where each vertex may be weighted according to the availability of a defense mitigation. Specifically, when $\beta > 1$, the selection of decoys exposing techniques with mitigations is discouraged.

Constraint \eqref{eq:reachable_or} and \eqref{eq:reachable_and} implement the reachability logic condition for each vertex based on its type. In detail, they ensure that each node is labeled as reachable from a source node in $S$, i.e. $r_{i,s} = 1$, only if its predecessors can be reached as well. More precisely, \eqref{eq:reachable_or} guarantees that a \textit{OR} vertex $v_i$ is reachable if it is assigned to the partition $Y$ or it is assigned to the $Z$ partition and at least one  of its predecessors, indicated by the set $N_{\bar{G}}(v_i)$, is reachable. Note that if $v_i$ is assigned to the partition $X$, it is unreachable by definition, hence $r_{i,s} = 0$ due to $y_i + z_i = 0$. Instead, \eqref{eq:reachable_and}
guarantees the reachability of a \textit{AND} vertex by assessing the opposite condition, i.e. when it is not reachable (we considered the unreachable condition for the nodes of type \textit{AND} as it allow us to employ the same variables without the need of introducing new ones). This condition occurs when a node of type AND belongs to $X$ partition and has at least one unreachable predecessor.
Constraint \eqref{eq:cross_boundaries} enforces that the nodes in the $X$ partition must disconnect any attack path between each $v_i \in Y$ and $v_j \in Z$ given the reachability of the various nodes. In particular, it forbids the existence of a directed edge $(v_i, v_j) \in \bar{E}$ such that $v_i \in Y$ and $v_j \in Z$ since this would violate the property of the $X$ partition unless either $v_i$ or $v_j$ is unreachable. 
Constraint \eqref{eq:init_var} initializes $Y$ and $Z$ with the vertexes in $S$ and $T$, respectively. Moreover, it ensures that $X$ does not contain any technique outcomes as they are not eligible to be selected as decoys.
Constraint \eqref{eq:one_cluster} guarantees that every vertex in the graph is assigned to one partition only. Finally, constraint \eqref{eq:binary} enforces the integer nature of the problem formulation.

The above formulation \eqref{eq:obj_func_new}-\eqref{eq:binary} has complexity NP-Hard as it can be reduced to the original Multi-Terminal Vertex Separator problem in polynomial time when nodes type is ignored. Note that the non-linearity introduced by the \textit{min} operator in constraints \eqref{eq:reachable_or}-\eqref{eq:reachable_and} can be linearized by using suitable auxiliary variables as explained in \citep{belov2016improved}.
In general, the non-polynomial complexity makes the solution computation unpractical for large graphs due to the high execution time. However, the number of nodes composing $\bar{G}$ is limited since it is constrained by the maximum number of available ATT\&CK techniques, hence setting an upper bound to the scaling complexity. Therefore, as we are going to show in the performance evaluation section, the design of a heuristic approximating the optimal solution is unnecessary due to the negligible resolution time that we experimentally observed during the simulations.

\subsection{Decoy selection solution overview}
We discuss an example of decoy selection solution to better showcase the various steps on a practical scenario and to provide a better understanding of the proposed methodology. We provide a visual representation of the solution in \cref{fig:mal_example}. \hl{We consider a defender that wants to allocate a number of decoys to expose a malicious user that has already a foothold within the environment and that aims at executing malicious code to infect a host within the network, as well as to steal user credentials}. The deployed decoys are selected according to the procedure described in the previous section. In detail, we identify the attacker target as the set $T = \{ infectedComputer, userCredential\}$, where \textit{infectedComputer} and \textit{userCredential} \hl{are technique outcomes indicating an attacker that has successfully run malicious software on a host and that has managed to steal account name and passwords of a legitimate user}. Similarly, we model the attacker initial foothold  by initializing the source node set as $S = \{userRights \}$, where \textit{userRights} is a technique outcome indicating an attacker that has acquired user right privileges and he is capable of interacting with most of the system assets. Given the defined $S$ and $T$, we can compute the threat profile $\bar{G}$ by first computing all the attack paths with \eqref{eq:ap_def} and pruning the unnecessary nodes and related edges from $G$. Then, we compute the minimum number of nodes that disconnect $S$ from $T$ solving  \eqref{eq:obj_func_new}-\eqref{eq:binary} \hl{under two different scenarios in order to better highlight the benefit of considering execution preconditions of type AND to reduce the number of decoys. Specifically, we compare the solution obtained when AND nodes are correctly considered with respect to a solution computed assuming that each node in $\bar{G}$ is of type OR only. Note that a solution which assumes each node to be of type AND would not ensure the interception of all attack paths. For example, a decoy exposing any of the execution preconditions needed to achieve an attack target of type OR  would be wrongly considered as optimal. For this reason we ignore this scenario as it would provide sub-optimal solutions that do not add any new meaningful insight to our approach. We visually represent the decoys selected by the two solutions in} \cref{fig:mal_example} \hl{ using yellow nodes for the “AND+OR solution” and purple nodes for “OR solution”, respectively. Note that double-colored nodes indicate that both solutions share the same decoy selection.}  

When AND and OR nodes are both considered, the optimal solution corresponds to three decoys exposing \textit{shortcutModification}, \textit{rightToLeftOverride} and \hl{\textit{exploitationForPrivilegeExalation}} techniques. In detail, \textit{shortcutModification} indicates an adversary modifying the shortcuts within the startup folder so that malicious code is executed upon system boot. A possible deceptive artifact may be a fake startup folder that treats any modification of its content as a malicious activity. The technique \textit{rightToLeftOverride} consists in the usage of the “right-to-left override” Unicode character to trick a legitimate user into opening malicious files. A possible deceptive artifact may be an agent that monitors files containing such character and that redirects their execution within a controlled and isolated environment to waste attacker resources in fake interactions. Finally, \textit{exploitationForPrivilegeExalation} \hl{denotes the utilization of any software vulnerabilities and/or system misconfiguration to gain administrator privileges. A possible deceptive artifact can consist of the deployment of an application with vulnerabilities that have been exploited to escalate privileges in order to lure and trap the attacker within an isolated environment. In contrast, when every node is assumed to be of type OR, the optimal solution selects five decoys (two more than the previous case) exposing the techniques \textit{exploitationForPrivilegeExalation}, \textit{indicatorRemoval},  \textit{shortcutModification}, \textit{spearphisingAttachment} and \textit{maliciousFile}.

The difference between the two solutions is explained by the role that AND nodes play in the solution computation. Intuitively, AND nodes are used by the optimal formulation to cleverly select OR nodes which can effectively minimize the number of required decoys. In the considered example, the selection of \textit{exploitationForPrivilegeExalation} and \textit{rightToLeftOverride} techniques computed by the “AND+OR solution" ensures the interception of any attack path containing \textit{clearEventLogs} and \textit{maliciousFile} techniques. As a matter of fact, the latter require AND preconditions, hence a proper choice of their predecessors can simultaneously intercept the related attack paths as well as any other attack paths composed by OR nodes. Conversely, the “OR solution", which ignores AND preconditions, must select multiple predecessors of AND nodes in order to prevent any possible attack paths, increasing the number of required decoys.  This behavior is further exacerbated in more complex threat profiles composed of multiple attack targets, hence an effective decoy selection exploiting the property of AND nodes can greatly reduce the number of needed decoys.}
Moreover, we further remark that this is an example meant to highlight the features of the proposed scheme in a simple scenario. The number of nodes composing the threat profile $\bar{G}$ largely depends on the number of considered source nodes and attack targets. More complex configurations generate threat profiles potentially composed by hundreds of nodes that offer less trivial decoy selection solutions and thus justify the need to provide an analytical problem formulation.

\section{Performance evaluation}\label{section:performance_evaluation}
\subsection{Simulation Setup}
The simulation environment was developed in Python. In particular, we used \textit{NetworkX} library to handle all the basic graph operations, whereas we implemented custom methods to handle the computation involving nodes of type \textit{AND}. We solved \eqref{eq:obj_func_new}-\eqref{eq:binary} using the \textit{SCIP} optimization suite, an open-source solver for integer linear programming \citep{scip}. We assessed the performance of the optimal decoy selection, referred to as \textbf{Optimal}, with the following baseline schemes: 
\begin{itemize}
    \item \textbf{Group}: this scheme selects decoys exposing the techniques used by one or more groups in the MITRE ATT\&CK framework. We recall that each group identifies the set of techniques used by real-world threat actors to reach some target. Therefore, the selected techniques constitute an attack path that was successfully employed  to accomplish some goal in a real enterprise system. As a result, differently from the optimal scheme that selects decoys by leveraging the proposed threat model to generate all feasible attack paths between a set of source and target nodes, this approach selects decoys tailored to intercept specific attack paths only. To ensure a fair performance comparison with optimal approach, we consider groups that are compatible with the attack targets in $\bar{G}$ by selecting those that employ techniques reaching the considered targets (in practice, we check if the predecessors nodes of the  targets nodes correspond to a technique used by the group). More precisely, in order to provide different group selection criteria, we extract groups from the pool provided by the MITRE ATT\&CK framework according to the parameters $\rho$ and $\gamma$ . The former is the minimum percentage of attack targets that a group must reach in order to be considered as compatible with the threat profile $\bar{G}$. The latter is the percentage of groups that the scheme randomly samples from the pool of compatible groups. For example, the decoys selected with the configuration $\gamma = 0.5$ and $\rho = 0.5$ expose the techniques used by the 50\% of groups chosen randomly from those compatible with at least the 50\% of the considered attack targets (note that, with abuse of notation, we indicate $\gamma = 0$ and $\rho = 0$ as the selection of one random group compatible with at least one attack target).
    \item \textbf{Predecessor}: this scheme selects the decoys that expose the techniques causing the accomplishment of each attack target in $T$. Intuitively, this approach can be considered as a graph partition heuristic that ignores the node type and that selects the decoys detecting all the preconditions associated to each target (in other words, these are the predecessor nodes of the target nodes in $\bar{G}$). As a result, it ensures the interception of every attack paths, since all the preconditions required to reach every target are detected by a corresponding number of decoys. 
     \item \textbf{Random}: this scheme selects decoys exposing a random subset of techniques sampled from the considered threat profile $\bar{G}$. The number of sampled techniques is the same as the one computed by the optimal scheme in order to equalize the performance assessment. In other words, this scheme should be intended as a performance lower bound and it is used to better highlight any gain provided by the optimal approach.
\end{itemize}

We compare the security performance of the proposed schemes using three main metrics that capture a different feature of the resulting decoy selection:
\begin{itemize}
\item \textbf{Attack path interception ratio}: this metric measures the deception performance of the selected decoys in luring attackers before they reach their goal once deployed. We compute its value as the ratio between the number of attack paths containing at least one technique that can be exposed by a decoy and the total number of attack paths in $\bar{G}$. A high value indicates an effective decoy placement that is tailored to the attacker behavior modeled using ATTA\&CK techniques.
\item \textbf{Number of decoys}: this metric measures the number of decoys selected by each scheme that are required to achieve the corresponding attack path interception performance. We use this metric to quantify the attack surface increase caused by the deployment of the various decoys. Therefore, assuming two decoy selection solutions with similar attack path interception performance, the one requiring the lowest amount of decoys is preferable to limit possible unpredicted impacts on the legitimate system components. 
\item \textbf{Unmitigated technique ratio}: this metric measures the ratio of  techniques exposed by the selected decoys that have no defense mitigation available. 
A high value of this metric suggests that the computed decoy selection can efficiently complement available security mechanisms for the considered threat profile $\bar{G}$ by ensuring the protection from techniques without available countermeasures.
\end{itemize}

\begin{figure*}
    \begin{minipage}[t]{.3\textwidth}
    \centering
    \includegraphics[width=1\textwidth]{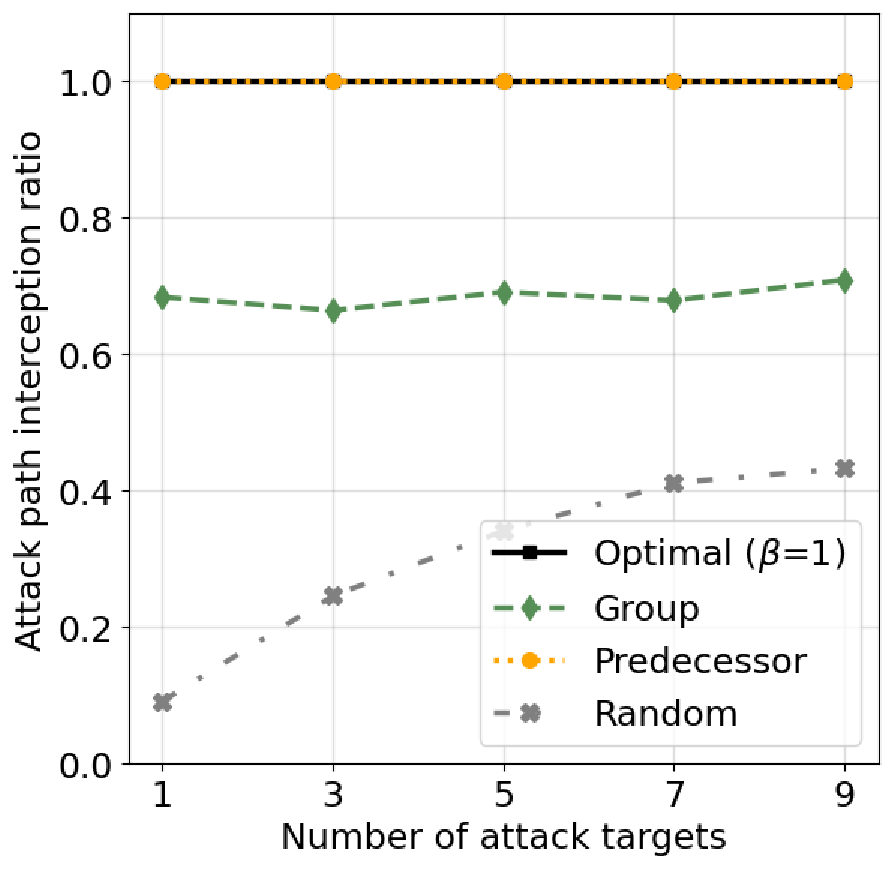}
    \caption{Ratio of attack paths intercepted by the decoys.}
    \label{fig:1_group_all_targets_paths}
    \end{minipage}
    \hfill
    \begin{minipage}[t]{.3\textwidth}
    \centering
    \includegraphics[width=1\textwidth]{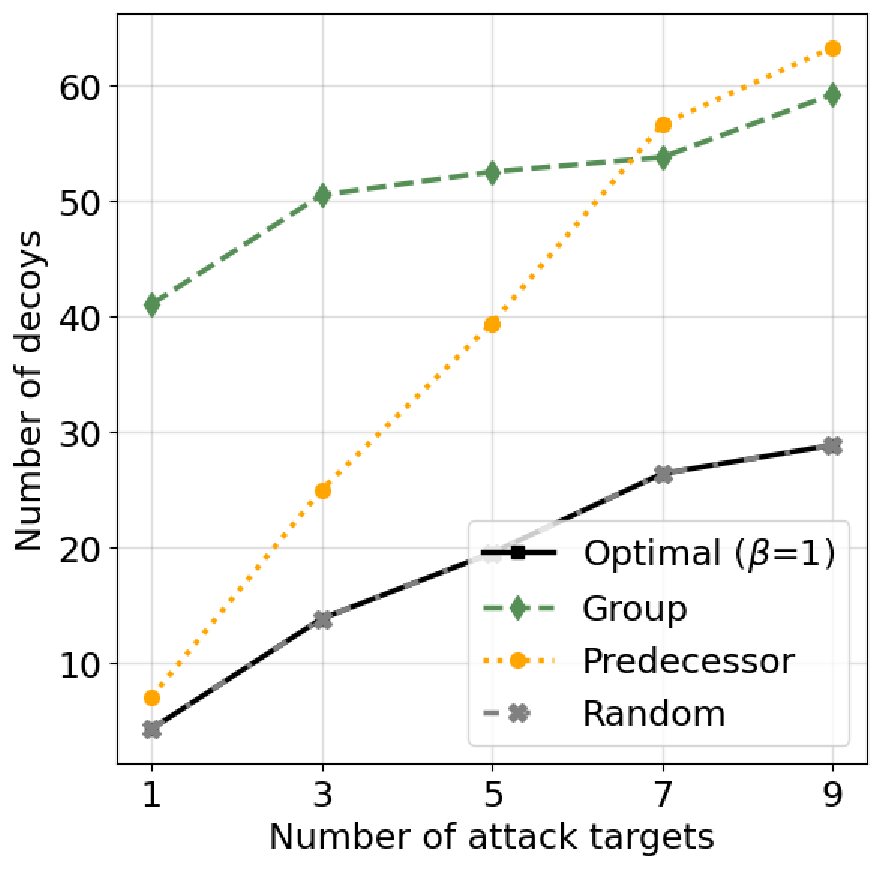}
    \caption{Number of selected decoys.}
    \label{fig:1_group_all_targets_num_decoys}
    \end{minipage}
    \hfill
    \begin{minipage}[t]{.3\textwidth}
    \centering
    \includegraphics[width=1\textwidth]{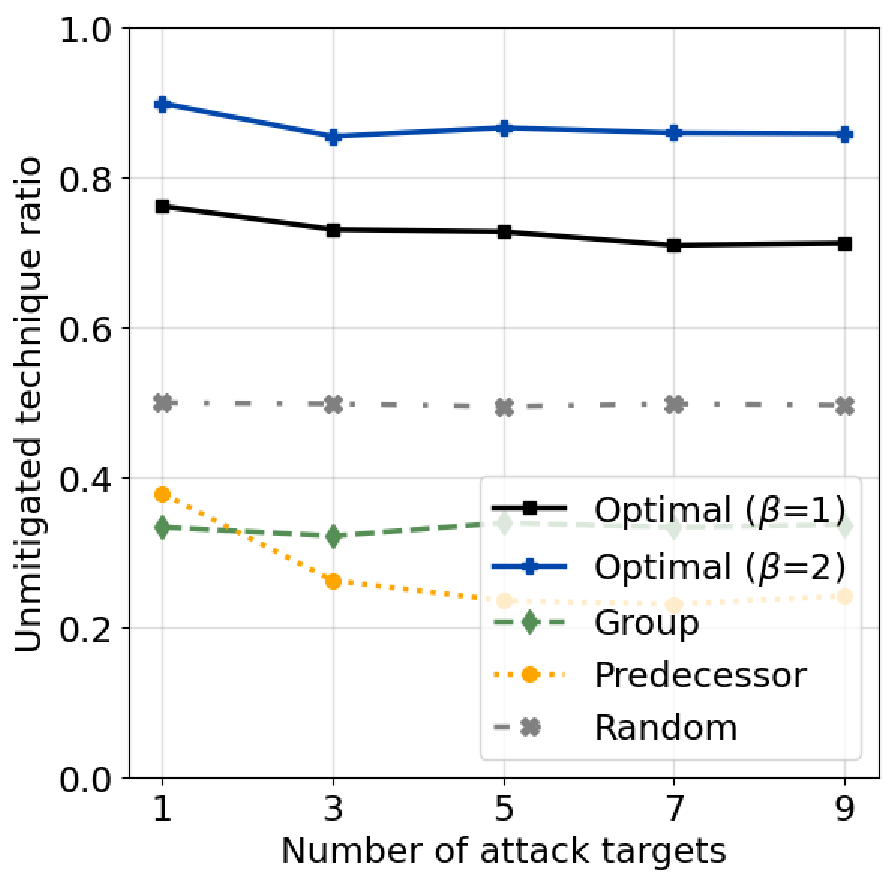}
    \caption{Ratio of techniques without mitigation exposed by the decoys.}
    \label{fig:1_group_all_targets_without_mit}
    \end{minipage}
\end{figure*}

We considered different configurations of attack targets ranging from 1 to 9 and averaged the results of 100 random instances of threat profiles $\bar{G}$ generated as follows. We modeled the initial attacker capabilities assuming $S = \{userRight \}$  as this reflects a realistic scenario where an attacker has gained user privileges within the system and leverages those privileges to further penetrate the environment with more advanced techniques. Although this is not comprehensive of every possible attack path, the majority of real-world attacks reported in the MITRE ATT\&CK framework shares this intermediate attack goal. As a matter of fact, a malicious user with no level of rights has a limited pool of available techniques which are indeed focused to gain some permission rights within the system, thus allowing to expand his capabilities. We selected  the various attack targets by randomly sampling a number of technique outcomes from the set $O$ for each configuration. Practically, we extracted the initial set of technique outcomes by individually inspecting each attack step encoded in enterpriseLang to check if it was composed by additional elements other than the technique name. This procedure produced 153 technique outcomes which were considered eligible to be selected as attack targets.

\subsection{Decoy selection performance analysis}
We evaluate the various schemes according to the aforementioned metrics. In particular, unless otherwise stated, the optimal scheme is configured with $\beta = 1$ (i.e., mitigation-unaware), the group scheme is computed with $\gamma = 0$ and $\rho = 1$ (i.e., one group whose techniques can reach all selected targets).

In \cref{fig:1_group_all_targets_paths}, we show the attack path detection ratio achieved by the various schemes according to a variable number of attack targets. The optimal scheme and the predecessor scheme achieve a 100\% interception rate of the attack paths for any number of attack targets. This is an expected behavior, as these methods are designed to compute a subset of nodes that disconnect any source in $S$ to any target in $T$. Therefore, the associated selected decoys intercept all attack paths. The performance gap between the optimal and group scheme is explained by the fact that the latter considers only techniques exclusively used by one threat actor and it ignores all possible attack paths enabled by other techniques composing the considered threat profile. Conversely, the optimal scheme is designed to comprehensively accounts for any feasible  attack path that allows to reach the various targets. The random scheme provides the worst performance due to the fact that it does not exploit any information provided by the attack graph, highlighting that the performance gain of optimal decoy selection leverages the preconditions shared among the employed techniques to efficiently deny multiple attack paths. Note that increasing performance gain of the random scheme is due to a higher likelihood to randomly intercept attack paths, which indeed increase with the number of targets.

In \cref{fig:1_group_all_targets_num_decoys}, we present the number of decoy selected by each scheme. As a general trend, the overall number of decoys increases for all schemes as a higher number of attack targets generates a higher number of attack paths that need to be intercepted. The optimal scheme always selects the lowest number of decoys, thus minimizing the impact on the attack surface of the system due to the deployed decoys. The random scheme also achieves the same performance since it is designed to select the same number of decoys as the optimal method. Conversely, the predecessor scheme provides a rapidly-increasing number of decoys as it restricted to select those techniques causing the outcomes chosen as targets. Therefore, the existence of multiple attack targets reachable by many techniques require a large amount of decoys. This behavior makes such scheme unpractical when the number of considered targets is high due to its poor scalability performance. Finally, the group scheme produces a considerable number of decoys even for a low number of targets but it less sensitive when this value increases. This depends on the flexibility of the techniques used by the associated threat actors as they are likely to be employed for a plethora of different attack scenarios, allowing their reusability for different attack targets. By combining the number of decoys with their attack path interception capability, we highlight the advantage provided by the optimal scheme with respect to the other approaches as it guarantees the highest interception rate using the lowest amount of decoys.

In \cref{fig:1_group_all_targets_without_mit}, we show the unmitigated technique ratio. The performance of the optimal method is dictated by the $\beta$ parameter, which associates higher weights to the nodes in $\bar{G}$ corresponding to techniques with mitigations. In particular, compared to the setting $\beta=1$ that provides a mitigation-unaware decoy selection, $\beta=2$ successfully selects a higher percentage of techniques without countermeasures to deny the various attack paths. Nonetheless, regardless of the value of  $\beta$, the optimal approach outperforms all other schemes by a significant margin, suggesting that the techniques allowing for a more efficient partition of $\bar{G}$ are indeed without mitigation and therefore the associated attack paths are more effectively detected via a proactive defense approach, such as cyber deception. This observation is further supported by the performance of the group and predecessor schemes which are instead biased to select a higher ratio of techniques with mitigation as decoys. Specifically, the group scheme selects techniques forming attack paths successfully executed in real-world systems, whereas the predecessor scheme selects techniques that have an immediate impact (i.e., the technique outcome) on the environment. In both scenarios, these techniques become ideal candidates for the design of possible defense countermeasures to mitigate them. As a result, the associated decoys are more likely to overlap with existing security mechanisms. Finally, the random scheme achieves constant performance regardless of the number of targets, as it is agnostic to the available mitigations.
 
\begin{table}[t]
    \centering
    \scriptsize
	\caption{Computational time (s)}
    \label{table:time}
    \vspace{-0.2cm}
    \begin{tabular}{ |c|c|c|c|c| }
        \hline
        Targets & Optimal & Predecessor & Group & Random \\
        \hline
        1 & $4.23 \cdot 10^{-2}$ & $1.91 \cdot 10^{-5}$ & $1.6 \cdot 10^{-4}$ & $6.27 \cdot 10^{-6}$ \\
        \hline
        5 & $4.90 \cdot 10^{-2}$ & $3.34 \cdot 10^{-5}$ & $1.72 \cdot 10^{-4}$ & $7.11 \cdot 10^{-6}$ \\
        \hline
        9 & $5.99 \cdot 10^{-2}$ & $4.97 \cdot 10^{-5}$ & $1.72 \cdot 10^{-4}$ & $8.07 \cdot 10^{-6}$ \\
        \hline
        
        \hline
    \end{tabular}
\end{table}

In \cref{table:time}, we report the computational time required by each scheme to calculate the decoy selection. As mentioned beforehand, despite the NP-Hard complexity of the optimal formulation, the limited dimension of the attack graph ensures a computational time performance for the optimal scheme that is negligible in practical scenarios and comparable to the one achieved by the other schemes. 

\subsection{Optimal Solution Analysis}
\begin{figure}
    \begin{minipage}[t]{.23\textwidth}
    \centering
    \includegraphics[width=1\textwidth]{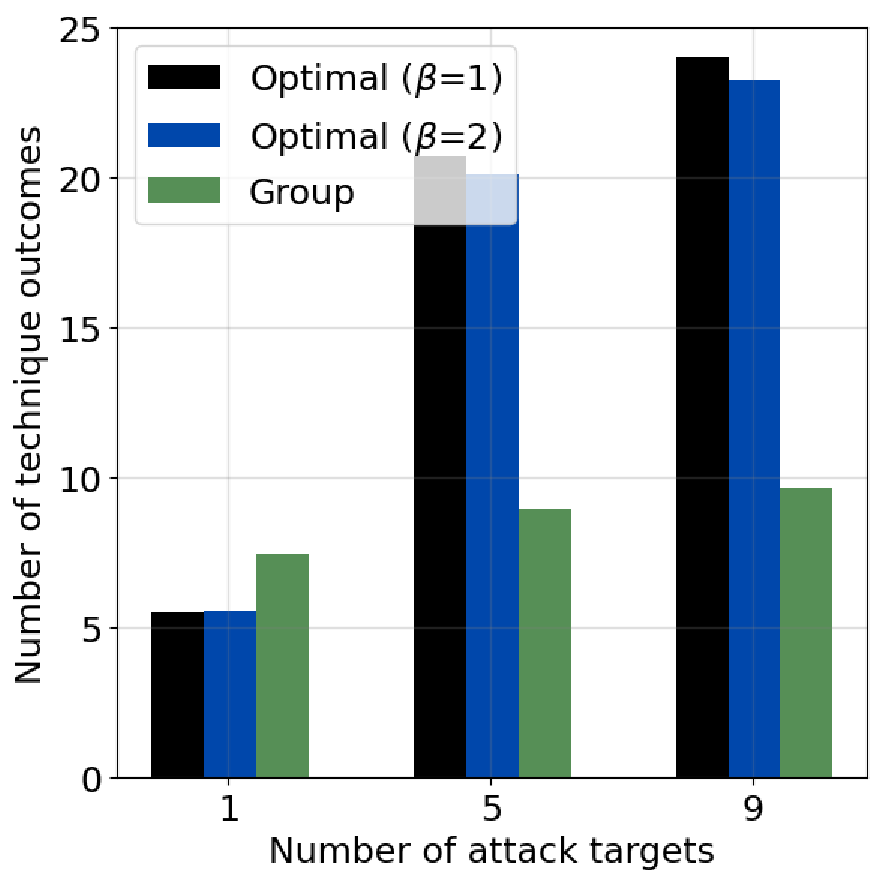}
    \caption{Number of prevented technique outcomes in addition to attack targets.}
    \label{fig:1_group_all_targets_not_reachable}
    \end{minipage}
    \hfill
    \begin{minipage}[t]{.23\textwidth}
    \centering
    \includegraphics[width=1\textwidth]{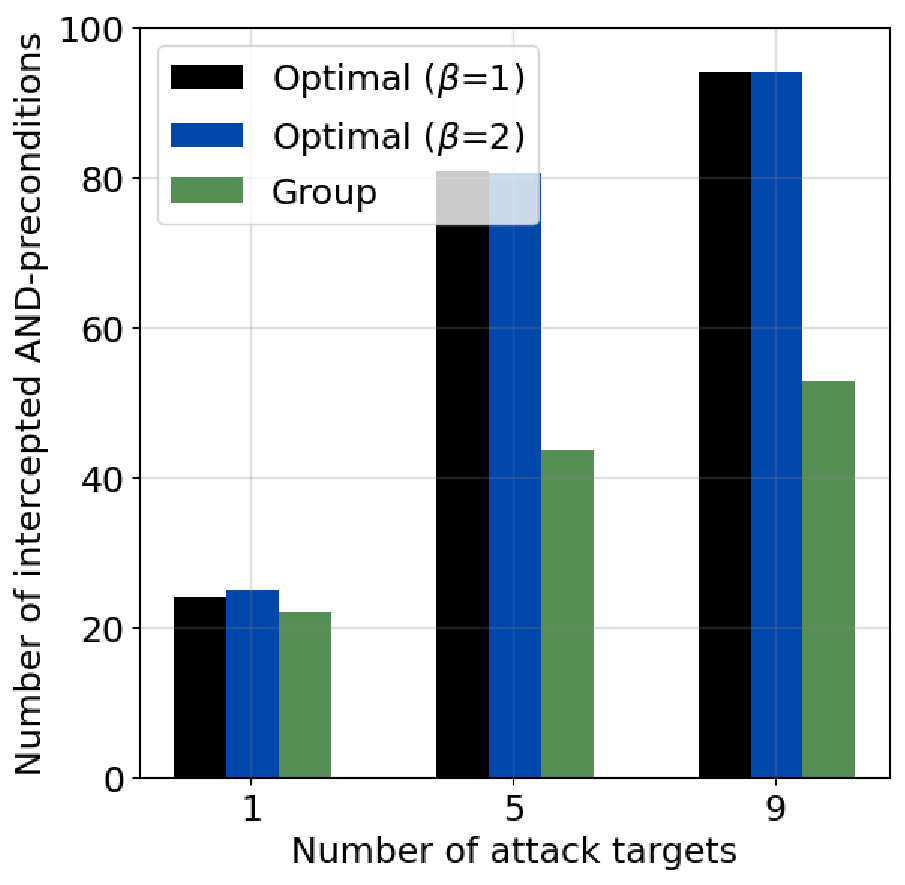}
    \caption{Number of AND nodes intercepted by the decoys.}
    \label{fig:1_group_all_targets_disconnected_and}
    \end{minipage}  
\end{figure}

We provide a deeper performance analysis of the optimal decoy selection solution to offer additional insights about its behavior compared to the group scheme. We exclude the  predecessor method since it can be considered as a low-complexity heuristic of the optimal scheme. 
In \cref{fig:1_group_all_targets_not_reachable}, we present the number of technique outcomes, not belonging to the set of attack targets, that cannot be reached by an attacker without interacting with a decoy. In other words, this metric evaluates the effectiveness of a decoy selection that is tailored to a specific set of attack targets in preventing attack paths directed to other targets not considered in the threat profile $\bar{G}$. The optimal solution ensures a higher number of “unintended” technique outcomes being detected by the decoys compared to the group scheme. This behavior suggests that the optimal scheme selects the techniques that are shared by many attack paths. From a topological standpoint, this means that it prioritizes nodes in $\bar{G}$ traversed by  multiple attack paths in order to reduce the number of required decoys. When the number of attack targets increase, this effect is further amplified as it can be observed by the increasing value of this metric. In contrast, the group scheme is moderately sensible to this effect despite using a considerable higher number of decoys as shown in the previous plot. This trend is consistent with its design approach that produces a decoy selection strictly tailored to the considered attack targets. Consequently, the decoys expose techniques shared by a fewer number of attack paths and hence they are not as effective as the optimal decoy selection in preventing attack paths to “unintended” technique outcomes. As a side note, the group slightly outperforms the optimal scheme when a single target is selected due to the huge difference in the number of selected decoys (i.e., 40 decoys for the group scheme vs. 4 decoys for the optimal scheme) that ensures more opportunities to prevent a higher amount of technique outcomes.

In \cref{fig:1_group_all_targets_disconnected_and}, we show the average number of AND nodes detected by the decoys. This metric quantifies the capability of the decoys computed with the optimal and group schemes to exploit AND preconditions in $\bar{G}$ in order to detect multiple attack paths. The optimal scheme achieves the best performance since it actively takes advantage of the AND logic that is embedded within the problem formulation. As a matter of fact, an attacker needs to satisfy every precondition of an AND node in order to employ the related technique or to achieve the resulting outcome. This dynamic allows reducing the number of required deceptive artifacts since any technique or outcome of type AND can be efficiently disabled by exposing one of its predecessor techniques with a decoy. Conversely, the group scheme is insensible to the node type as it is shown by the evident performance gap.

\subsection{Group Selection Sensitivity}
We provide a more extensive comparison between the optimal scheme and group scheme by generating different combinations of $\gamma$ and $\rho$ parameters as reported in \cref{tab:group_conf}.  

\begin{table}[t]
    \centering
    \scriptsize
	\caption{Group scheme configuration}
     \label{tab:group_conf}
    \vspace{-0.2cm}
    \begin{tabular}{ |l|l|l|  }
        \hline
        $\gamma$ & $\rho$ & Configuration\\
        \hline
        $0$ & $0$ & Single threat actor reaching at least one attack target \\
        \hline
        $0$ & $1$ & Single threat actor reaching all attack targets \\
        \hline
        $1$ & $0$ & All threat actors reaching at least one attack target \\
        \hline
        $1$ & $1$ & All threat actors reaching all attack targets \\
        \hline
    \end{tabular}
\end{table}

\begin{figure}[t]
    \begin{minipage}[t]{.23\textwidth}
    \centering
    \includegraphics[width=1\textwidth]{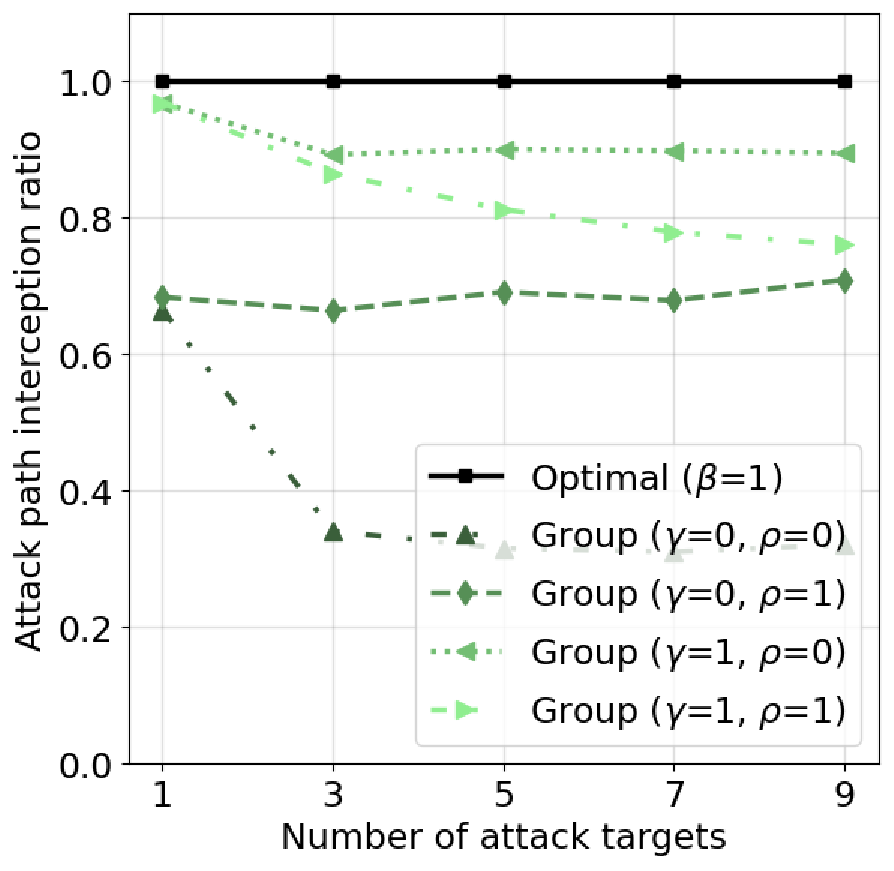}
    \caption{Ratio of attack paths intercepted using different group scheme configurations.}
    \label{fig:group_selection_paths}    
    \end{minipage}
    \hfill
    \begin{minipage}[t]{.23\textwidth}
    \centering
    \includegraphics[width=1\textwidth]{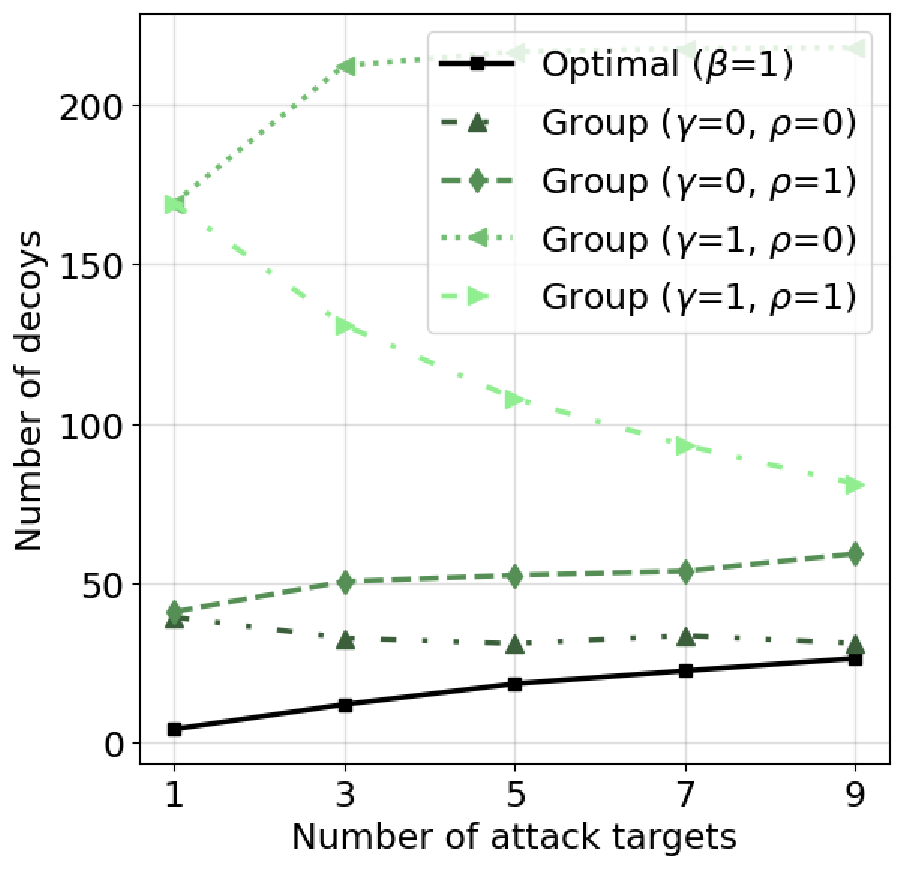}
    \caption{Number of selected decoys using different group scheme configurations.}
    \label{fig:group_selection_n_decoys}
    \end{minipage}
\end{figure}

In \cref{fig:group_selection_paths}, we show the attack path interception ratio. In general, the group scheme configurations selecting all threat actors (i.e., $\gamma = 1$)  detects a higher number of attack paths. These configurations select a higher number of decoys compared to the single threat actor approach (i.e., $\gamma = 0$) as they select a higher amount of techniques compatible with the attack targets. 
Similarly, the compatibility between threat actors and attack targets also impacts the attack path interception performance of the group scheme. The selection of threat actors reaching every attack target (i.e., $\rho = 1$) tends to ensure a higher ratio of intercepted attack paths since the selected decoys are more likely to detect techniques composing the considered threat profile. However, when the number of target increases, the selection of all threat actors reaching at least one attack target (i.e., $\gamma = 1,\rho = 0$) ensures a higher ratio of intercepted attack paths compared to selecting threat actors reaching all attack targets (i.e., $\gamma = 1,\rho = 1$). This counterintuitive behavior is due to the limited number of groups satisfying the target reachability condition. As the number of attack targets increases, there are fewer threat actors that can simultaneously reach each target. This filtering reduces the number of selected decoys since the number of technique to expose is progressively lower. Consequently, the selected decoys are strictly tailored to detect attack paths used by few threat actors only. In other words, this configuration approximates the single-threat actor approach for a high number of targets as the pool of eligible threat actors is composed by few units (i.e., similarly to the configuration $\gamma = 1$), thus explaining the comparable performance. 
Conversely, when the reachability requirement is $\rho = 1$, we observe a more stable performance due to the fact that the scheme has a larger pool of threat actors to choose from, which heavily inflates the attack path detection performance due to the large number of selected decoys. However, when combined with the single threat actor configuration, the resulting decoy selection yields the worst results since the limited number of exposed techniques are rarely used by the various attack paths to reach the considered targets.

In \cref{fig:group_selection_n_decoys}, we show the number of selected decoys. In general, the group scheme configurations employing multiple threat actors also select a higher number of decoys. This is consistent with their related attack path detection performance, which is inflated by the number of the available techniques that can be exposed by a suitable decoy. This trend further demonstrates the advantage of the optimal scheme, whose attack path detection performance derives from the exploitation of the execution preconditions between various techniques to limit the decoys number. Conversely, for the opposite reason, the group configurations using single threat actors employ a lower amount of decoys. As mentioned beforehand, we remark that the decreasing behavior of the configuration $\gamma = 1$ and $\rho = 1$ is due to the limited number of threat actors satisfying the sampling conditions as the number of targets increases, which in turn reduces the number of available decoys.

The performance trade-off across the analyzed metrics shows that the configuration  $\gamma = 1$ and $\rho = 0$ achieve the most balanced decoy selection results in terms of attack path interception ratio and number of deceptive artifacts. Nonetheless, it is outperformed by the proposed optimal scheme due to its fundamentally different design approach that considers the cause-effect correlation of the techniques to boost the deception performance while causing a modest attack surface increase.

\section{Conclusion}\label{section:conclusion}
We proposed a decoy selection scheme to proactively select deceptive artifacts using an adversarial model built from empirical observations of TTPs. Specifically, we first modeled the attacker capabilities by combining the information provided by enterpriseLang, a threat language modelling for enterprise environments, and MITRE ATT\&CK framework. The resulting threat model allows to generate attack graphs expressing the various attack paths between a source node, representing the initial attacker foothold, and a set of attack targets, representing the final attacker goals, using real-world attackers techniques. Each attack path is expressed as a subset of ATT\&CK techniques that can be exposed by suitable decoys designed to lure attackers employing those strategies. Then, based on this assumption, we formulated an integer linear optimization problem that computes the minimum number of decoys exposing ATT\&CK techniques required to reach a set of predefined attack targets. The proposed formulation accounts for the execution precondition of the various techniques to properly optimize the amount of deceptive artifacts. As a result, the selected decoys can effectively intercept attackers before they accomplish those goals. We assessed the performance of our approach by designing different schemes that partially exploit the cyber threat information provided by our model to compute the decoy selection. The proposed optimal scheme outperformed all the considered benchmarks in all metrics. In particular, by taking advantage of logical dependencies between the various techniques, it selects a considerably lower amount of decoys to intercept all attack paths directed to a specific set of attack targets.

\bibliographystyle{elsarticle-harv} 
\bibliography{bibliography}

\end{document}